\documentclass{article}
\usepackage{amssymb}
\usepackage{graphicx}
\usepackage{amsmath}

\input{tcilatex}

\begin{document}

\title{Spin-1 gravitational waves. \\
Theoretical and experimental aspects\thanks{%
The article is dedicated, in the occasion of his 60th birthday, to Raphael
Sorkin who first introduced gravitational particles with spin different from
2.}}
\author{F. Canfora, L. Parisi, G. Vilasi \\
\textit{Dipartimento di Fisica E.R.Caianiello, Universit\`{a} di Salerno.}\\
\textit{Istituto Nazionale di Fisica Nucleare, GC di Salerno, Italy.}\\
}
\maketitle

\begin{abstract}
Exact solutions of Einstein field equations invariant for a non-Abelian
2-dimensional Lie algebra of Killing fields are described. Physical
properties of these gravitational fields are studied, their wave character
is checked by making use of covariant criteria and the observable effects of
such waves are outlined. The possibility of detection of these waves with
modern detectors, spherical resonant antennas in particular, is sketched.
\end{abstract}

\section*{Introduction}

Gravitational waves, that is a propagating warpage of space time generated
from compact concentrations of energy, like neutron stars and black holes,
have not yet been detected directly, although their indirect influence has
been seen and measured with great accuracy. Presently there are, worldwide,
many efforts to detect gravitational radiation, not only because a direct
confirmation of their existence is interesting \textit{per se} but also
because new insights on the nature of gravity and of the Universe itself
could be gained. For these reasons exact solutions of the Einstein field
equations deserve special attention when they are of propagative nature. The
need of taking into full account the nonlinearity of Einstein's equations
when studying the generation of gravitational waves from strong sources is
generally recognized \cite{Th95}. Moreover, despite the great distance of
the sources from Earth (where most of the experimental devices, laser
interferometers and resonant antennas, are located) there are situations
where the non linear effects cannot be neglected. This is the case when the
source is a binary coalescence: indeed it has been shown \cite{Ch91} that a
secondary wave, called the \textit{Christodoulou memory} is generated via
the non linearity of Einstein's field equations. The memory seems to be too
weak to be detected from the present generation of interferometers \cite%
{Th92} (even if its frequency is in the optimal band for the LIGO/VIRGO
interferometers) but of the same order as the linear effects related to the
same source, thus stressing the relevance of the nonlinearity of the
Einstein's equations also (soon) from an experimental point of view.

On the theoretical side, starting from the seventy's new powerful
mathematical methods have been developed to deal with nonlinear evolution
equations. For instance, a suitable generalization of the \textit{\ Inverse
Scattering Transform }allows to integrate \cite{BZ78} Einstein field
equations for a metric of the form 
\begin{eqnarray*}
g=f\left( z,t\right) \left( dt^{2}-dz^{2}\right) +h_{11}\left( z,t\right)
dx^{2}+h_{22}\left( z,t\right) dy^{2}+2h_{12}\left( z,t\right) dxdy.
\end{eqnarray*}%
Indeed, the corresponding vacuum Einstein field equations reduce essentially%
\footnote{%
The function $f$ can be obtained by quadratures in terms of the matrix $%
\mathbf{\ \ H}$.} to 
\begin{eqnarray*}
\left( \alpha \mathbf{H}^{-1}\mathbf{H,}_{\xi }\right) ,_{\eta }+\left(
\alpha \mathbf{H}^{-1}\mathbf{H,}_{\eta }\right) ,_{\xi }=0,
\end{eqnarray*}%
where $\mathbf{H}\equiv $ $\left\Vert h_{ab}\right\Vert ,$ $~\xi =\left(
t+z\right) /\sqrt{2},\quad \eta =\left( t-z\right) /\sqrt{2},\quad \alpha =%
\sqrt{\left\vert \det \mathbf{H}\right\vert }$. This is a non-linear
differential equation whose form is typical of two-dimensional integrable
systems and can be integrated \cite{BZ78} by using a suitable generalization
of the \textit{\ Inverse Scattering Transform}, yielding \textit{\ solitary
waves solutions}.

A geometric inspection of the metric above shows that it is invariant under
translations along the $x,y$-axes, \textit{i.e. }it admits two Killing
fields, $\partial _{x}$ and $\partial _{y}$, closing on an Abelian
two-dimensional Lie algebra $\mathcal{A}_{2}$. Moreover, the distribution $%
\mathcal{D}$, generated by ${\partial _{x}}$ and ${\partial _{y}}$, is $2$%
-dimensional and the distribution $\mathcal{D}^{\mathcal{\perp }}$
orthogonal to $\mathcal{D}$ is integrable and transversal to $\mathcal{D}$.

Thus, it has been natural to consider \cite{SVV01} the general problem of
characterizing all gravitational fields $g$ admitting a Lie algebra $%
\mathcal{G}$ of Killing fields such that:

\begin{itemize}
\item[I] the distribution $\mathcal{D}$, generated by vector fields of $%
\mathcal{G}$, is $2$-dimensional;

\item[II] the distribution $\mathcal{D}^{\mathcal{\perp }}$, orthogonal to $%
\mathcal{D}$ is integrable and transversal to $\mathcal{D}.$
\end{itemize}

According to whether $\dim \mathcal{G}$ is $2$ or $3$, two qualitatively
different cases occur.

A 2-dimensional $\mathcal{G}$, is either Abelian ($\mathcal{A}_{2}$) or
non-Abelian ($\mathcal{G}_{2}$). A metric $g$ satisfying I and II, with $%
\mathcal{\ G}=\mathcal{A}_{2}$ or $\mathcal{G}_{2}$, will be called $%
\mathcal{G}$\textit{\ -integrable}.

The study of $\mathcal{A}_{2}$-integrable Einstein metrics goes back to
Einstein and Rosen \cite{ER37}, Rosen \cite{Ro54}, Kompaneyets \cite{Ko58},
Geroch \cite{Ge72}, Belinsky and Khalatnikov \cite{BK70}. Recent results can
be found in \cite{CFV04b}

The greater rigidity of $\mathcal{G}_{2}$-integrable metrics, for which some
partial results can be found in \cite{Ha88, AL92, Ch98}, allows an
exhaustive analysis. It will be shown that they are parameterized by
solutions of a linear second order differential equation on the plane which,
in its turn, depends linearly on the choice of a $\mathbf{j}$\textit{%
-harmonic} function (see later). Thus, this class of solutions has a \textit{%
bilinear structure} and, as such, admits two\textit{\ superposition laws. }

When $\dim $ $\mathcal{G}=3$, assumption II follows from I and the local
structure of this class of Einstein metrics can be explicitly described.
Some well known exact solutions \cite{Pe69,SKMHH03}, \textit{e.g.}
Schwarzschild, belong to this class.

Besides the new local $\mathcal{G}_{2}$-integrable solutions, a procedure to
construct new global solutions, suitable for all such $\mathcal{G}$%
-integrable metrics, will be also described.

The paper is organized as follows. In section \textbf{1} gravitational
fields invariant for a two dimensional Lie algebra are characterized. In
section \textbf{2} the Einstein equations for such metrics are reduced, by
using the symmetry, to the so called $\mu $-deformed Laplace equation.
Harmonic coordinates are also introduced. Section \textbf{3} is devoted to
the analysis of the wave-like character of the solutions through the
Zel'manov and the Pirani criterions. In section \textbf{4} the canonical and
the Landau energy-momentum pseudo-tensors are introduced and a comparison
with the linearised theory is performed. In section \textbf{5} realistic
sources for such gravitational fields are described. Eventually, section 
\textbf{6} is devoted to the analysis of the polarization of the waves.

In the following, $\mathcal{K}il\left( g\right) $ will denote the Lie
algebra of all Killing fields of a metric $g$ while \textit{Killing algebra}
will denote a sub-algebra of $\mathcal{K}il\left( g\right) $.

Moreover, an integral (two-dimensional) submanifold of $\mathcal{D}$ will be
called a \textit{Killing leaf}, and an integral ($2$-dimensional)
submanifold of $\mathcal{D}^{\bot }$ \textit{orthogonal leaf .}

\section{Geometrical aspects}

\begin{itemize}
\item \textit{Semiadapted coordinates.}

Let $g$ be a metric on the space-time $M$ (a connected smooth manifold) and $%
\mathcal{G}_{2}$ one of its Killing algebras whose generators $X,Y$ satisfy $%
[X,Y]=sY,\ \ \ s=0,1$

The Frobenius distribution $\mathcal{D}$ generated by $\mathcal{G}_{2}$ is
2-dimensional and a chart $(x^{1},\,x^{2},\,x^{3},\,x^{4})$ exists such that 
\begin{eqnarray*}
X=\frac{\partial }{\partial x^{3}},\,\,\ \,Y=\left( \exp sx^{3}\right) \frac{%
\partial }{\partial x^{4}}
\end{eqnarray*}%
From now on such a chart will be called \textit{semiadapted }(to the Killing
fields).

\item \textit{Invariant metrics}

It can be easily verified \cite{SVV01, SVV02a, SVV02b} that in a semiadapted
chart $g$ has the form%
\begin{eqnarray*}
g &=&g_{ij}dx^{i}dx^{j}+2\left( l_{i}+sm_{i}x^{4}\right)
\,dx^{i}dx^{3}-2m_{i}dx^{i}dx^{4}+ \\
&&\left( s^{2}\lambda \left( x^{4}\right) ^{2}-2s\mu x^{4}+\nu \right)
dx^{3}dx^{3}+ \\
&&2\left( \mu -s\lambda x^{4}\right) dx^{3}dx^{4}+\lambda dx^{4}dx^{4},\quad
i=1,2;j=1,2
\end{eqnarray*}

with\ $g_{ij}$, $m_{i}$, $l_{i}$, $\lambda $, $\mu $, $\nu $ arbitrary
functions of $\left( x^{1},x^{2}\right) $.

\item \textit{Killing leaves.}

Condition II allows to construct semi-adapted charts, with new coordinates $%
\left( x,y,x^{3},x^{4}\right) $, such that the fields $e_{1}=\partial
/\partial x,$ $e_{2}=\partial /\partial y$, belong to $\mathcal{D}^{\bot }$.
In such a chart, called from now on \textit{adapted}, the components $l_{i}$%
's and $m_{i}$'s vanish.

We will call \textit{Killing leaf} an integral (2-dimensional) submanifold
of $\mathcal{D}$ and \textit{orthogonal leaf } an integral (2-dimensional)
submanifold of $\mathcal{D}^{\bot }$. Since $\mathcal{D}^{\bot }$ is
transversal to $\mathcal{D}$, the restriction of $g$ to any Killing leaf, $S$%
, is non-degenerate. Thus, $\left( S,\left. g\right\vert _{S}\right) $ is a
homogeneous 2-dimensional Riemannian manifold. Then, the Gauss curvature $%
K\left( S\right) $ of the Killing leaves is constant (depending on the
leave). In the chart ($p=\left. x^{3}\right\vert _{S}$, $q=\left.
x^{4}\right\vert _{S}$) one has 
\begin{eqnarray*}
\left. g\right\vert _{S}=\left( s^{2}\widetilde{\lambda }q^{2}-2s\widetilde{%
\mu }q+\widetilde{\nu }\right) dp^{2}+2\left( \widetilde{\mu }-s\widetilde{%
\lambda }q\right) dpdq+\widetilde{\lambda }dq^{2},
\end{eqnarray*}%
where $\widetilde{\lambda },\widetilde{\mu },\widetilde{\nu }$, being the
restrictions to $S$ of ${\lambda },{\mu },\nu $, are constants, and 
\begin{equation}
K\left( S\right) =\widetilde{\lambda }s^{2}\left( \widetilde{\mu }^{2}-%
\widetilde{\lambda }\widetilde{\nu }\right) ^{-1}.  \notag
\end{equation}
\end{itemize}

\subsection{\textbf{Einstein metrics when $g(Y,Y)\neq 0$.}}

In the considered class of metrics, vacuum Einstein equations, $R_{\mu \nu
}=0$, can be completely solved \cite{SVV01}. If the Killing field $Y$ is not
of \textit{light type}, \textit{i.e.} $g(Y,Y)\neq 0$, then in the adapted
coordinates $\left( x,y,p,q\right) $ the general solution is 
\begin{equation}
g=f(dx^{2}\pm dy^{2})+\beta
^{2}[(s^{2}k^{2}q^{2}-2slq+m)dp^{2}+2(l-skq)dpdq+kdq^{2}]  \label{snew}
\end{equation}%
where $f=-\bigtriangleup _{\pm }\beta ^{2}/2s^{2}k$, and$\ \ \beta \left(
x,y\right) $ is a solution of the \textit{tortoise equation} 
\begin{eqnarray*}
\beta +A\ln \left\vert \beta -A\right\vert =u\left( x,y\right) ,
\end{eqnarray*}%
where $A$ is a constant and the function $u$ is a solution either of Laplace
or d' Alembert equation, $\bigtriangleup _{\pm }u=0$, $\bigtriangleup _{\pm
}=\partial _{xx}^{2}\pm \partial _{yy}^{2}$, such that $\left( \partial
_{x}u\right) ^{2}\pm \left( \partial _{y}u\right) ^{2}\neq 0$. The constants 
$k,l,m$ are constrained by $km-l^{2}=\mp 1,k\neq 0$ for Lorentzian metrics
or by $km-l^{2}=\pm 1,k\neq 0$ for Kleinian metrics (if $f>0$). Ricci flat
manifolds of Kleinian signature appear in the \textit{no boundary} proposal
of Hartle and Hawking \cite{HH83} in which the idea is suggested that the
signature of the space-time metric may have changed in the early universe.
Some other examples of Kleinian geometry in physics occur in the theory of
heterotic $N=2$ string (see \cite{OV91} and \cite{BGPPR94}) for which the
target space is four dimensional

\subsubsection{Canonical form of metrics when $g(Y,Y)\neq 0$}

The gauge freedom of the above solution, allowed by the function $u,$ can be
locally eliminated by introducing the coordinates $(u,v,p,q)$, the function $%
v(x,y)$ being conjugate to $u(x,y)$, \textit{i.e.} $\bigtriangleup _{\pm
}v=0 $ and $u_{x}=v_{y},u_{y}=\mp v_{x}$. In these coordinates the metric $g$
takes the form (local ''Birkhoff's theorem'')

\begin{eqnarray*}
g=\frac{\exp {\frac{u-\beta }{A}}}{2s^{2}k\beta }(du^{2}\pm dv^{2})+\beta
^{2}[(s^{2}k^{2}q^{2}-2slq+m)dp^{2}+2(l-skq)dpdq+kdq^{2}]
\end{eqnarray*}
with $\beta \left( u\right) $ a solution of $\beta +A\ln \left\vert \beta
-A\right\vert =u$.

\subsubsection{Normal form of metrics when $g(Y,Y)\neq 0$.}

In \textit{geographic coordinates} $\left( \vartheta ,\varphi \right) $
along Killing leaves one has 
\begin{eqnarray*}
\left. g\right\vert _{S}=\beta ^{2}\left[ d\vartheta ^{2}+\mathcal{F}\left(
\vartheta \right) d\varphi ^{2}\right] ,
\end{eqnarray*}
where $\mathcal{F}\left( \vartheta \right) $ is equal either to $\sinh
^{2}\vartheta $ or $-\cosh ^{2}\vartheta $, depending on the signature of
the metric. Thus, in the \textit{normal coordinates}, $\left( r=2s^{2}k\beta
,\tau =v,\vartheta ,\varphi \right) $, the metric takes the form 
\begin{equation}
g=\varepsilon _{1}\left( \left[ 1-\frac{A}{r}\right] d\tau ^{2}\pm \left[ 1-%
\frac{A}{r}\right] ^{-1}dr^{2}\right) +\varepsilon _{2}r^{2}\left[
d\vartheta ^{2}+\mathcal{F}\left( \vartheta \right) d\varphi ^{2}\right]
\label{nfs}
\end{equation}
where $\varepsilon _{1}=\pm 1$, $\varepsilon _{2}=\pm 1$ with a choice
coherent with the required \textit{signature} $2$.

The geometric reason for this form is that, when $g(Y,Y)\neq 0$, a third
Killing field exists which together with $X$ and $Y$ constitute a basis of $%
so(2,1)$. The larger symmetry implies that the geodesic equations describe a 
\textit{non-commutatively integrable system } \cite{SV00}, and the
corresponding geodesic flow projects on the geodesic flow of the metric
restricted to the Killing leaves. \textit{The above local form does not
allow, however, to treat properly the singularities appearing inevitably in
global solutions.}\emph{\ }The metrics (\ref{snew}), although they all are
locally diffeomorphic to (\ref{nfs}), play a relevant role in the
construction of new global solutions as described later.

\subsection{\textbf{Einstein metrics when $g(Y,Y)=0$.}}

If the Killing field $Y$ is of \textit{\ \ light type}, then the general
Lorentzian solution of vacuum Einstein equations, in the adapted coordinates 
$\left( x,y,p,q\right) $, is given by 
\begin{equation}
g=2f(dx^{2}+dy^{2})+\mu \lbrack (w\left( x,y\right) -2sq)dp^{2}+2dpdq]\text{,%
}  \label{gm}
\end{equation}%
where $\mu =A\Phi +B$ with $A,B\in \mathfrak{%
\mathbb{R}
}$, $\Phi $ is a non constant harmonic function of $x$ and $y$, $f=\left(
\nabla \Phi \right) ^{2}\sqrt{\left\vert \mu \right\vert }/\mu $, and $%
w\left( x,y\right) $ is solution of the $\mu $\emph{-deformed Laplace
equation}\textit{:} 
\begin{eqnarray*}
\mu \Delta w+\nabla \mu \cdot \nabla w=0.
\end{eqnarray*}

Metrics (\ref{gm}) are Lorentzian if the orthogonal leaves are conformally
Euclidean, \textit{i.e}. the positive sign is chosen, and Kleinian if not.
Only the Lorentzian case will be analyzed and these metrics will be called
of $\left( \mathcal{G}_{2},2\right) -$\textit{isotropic type}

In the particular case $s=1$, $f=1/2$ and $\mu =1$, the above metrics are
locally diffeomorphic to a subclass of the vacuum Peres solutions \cite{Pe59}%
, that for later purpose we rewrite in the form 
\begin{equation}
g=dx^{2}+dy^{2}+2dudv+2(\varphi _{,x}dx+\varphi _{,y}dy)du,  \label{spin1}
\end{equation}%
where 
\begin{eqnarray*}
u=e^{p},~~~~v=qe^{-p}+\varphi \left( x,y,u\right) ,
\end{eqnarray*}%
with $\varphi (x,y,u)$ a harmonic function of $x$ and $y$ arbitrarily
dependent on $u$.

\smallskip In the case $\mu =$ $const$, the $\mu $-deformed Laplace equation
reduces to the Laplace equation; for $\mu =1,$ in the harmonic coordinates
system $\left( x,y,z,t\right) $:%
\begin{eqnarray*}
z &=&\left[ \left( 2q-w\left( x,y\right) \right) e^{-p}+e^{p}\right] /2 \\
t &=&\left[ \left( 2q-w\left( x,y\right) \right) e^{-p}-e^{p}\right] /2,
\end{eqnarray*}%
the Einstein metrics (\ref{gm}) take \cite{CVV02} the particularly simple
form , 
\begin{equation}
g=2f(dx^{2}+dy^{2})+dz^{2}-dt^{2}+d\left( w\right) d\left( \ln \left\vert
z-t\right\vert ]\right) .  \label{gw}
\end{equation}%
This shows that, when $w$ is constant, the Einstein metrics given by Eq. (%
\ref{gw}) are static and, under the further assumption $\Phi =x\sqrt{2}$,
they reduce to the Minkowski one. Moreover, when $w$ is not constant,
gravitational fields (\ref{gw}) look like a \textit{disturbance} propagating
at light velocity along the$\ z$ direction on the Killing leaves (integral
two-dimensional submanifolds of $\mathcal{D}$).

\section{Physical properties}

The wave character of gravitational fields (\ref{gm}) has been checked by
using covariant criteria. In the following we will shortly review the most
important properties of these waves.

\subsection{Asymptotic flatness}

A first step toward a physical interpretation of metrics (\ref{gw}) is to
characterize those which are spatially asymptotically flat. For the metrics (%
\ref{gw}), in the vacuum case, we will consider (spatially) \textit{%
asymptotic} \textit{flat} a metric approaching the Minkowski metric for $%
x^{2}+y^{2}\rightarrow \infty $ \textit{.} In terms of the functions $f,$ $%
\mu $ and $w$, this asymptotic flatness condition reads: 
\begin{eqnarray*}
x^{2}+y^{2}\rightarrow \infty \implies f\rightarrow const,\quad \mathbb{\mu }%
\rightarrow const,\quad w\rightarrow c_{1}x+c_{2}y+c_{3}\text{,}
\end{eqnarray*}%
where $c_{1},c_{2}$ and $c_{3}$ are arbitrary constants and the behavior of $%
w$ can be easily recognized by looking at the Riemann tensor. In order for
metrics (\ref{gm}) to be spatially asymptotically flat $\mu $ must be
constant \cite{CVV02}. For $\mu =1$, the equation for $w$ reduces to a
two-dimensional Laplace equation, then in the vacuum case the only possible
choice is $w=const$, because the Laplace equation does not have solutions
tending to a constant value.

Let's consider the non-vacuum case. The simplest source for metrics (\ref{gm}%
) is \textit{dust} with density $\rho $ and velocity\textit{\ }$U^{\mu }$%
\textit{\ }with an energy-momentum tensor $T_{\mu \nu }=\rho U_{\mu }U_{\nu
} $ \cite{CV04}. When $U^{\mu }$ is a light-like vector field, this tensor
can describe the energy and momentum of \textit{null electromagnetic waves}.
This is not surprising; in fact Peres himself \cite{Pe59,Pe60} indicated
this as a possible source for his metrics (PP-waves) which, as we know, are
diffeomorphic to a subclass of solutions found in \cite{SVV01, SVV02a,
SVV02b}.\ With this stress-energy tensor we can depict realistic
astrophysical sources as \emph{Gamma Ray Bursts} or \emph{Cosmic Strings%
\footnote{%
Possible observations of Cosmic Strings have been reported recently \cite%
{Sa03}.}}; the symmetries of the vacuum solution is preserved.

Being the \textit{time coordinate} in the Killing leaves, the dust will be
chosen to move parallel to the \textit{light-like} Killing field $Y$, 
\textit{i.e., }with \textit{velocity} $U^{\mu }=\delta ^{\mu q}$. The non
vacuum Einstein equations with the energy-momentum tensor 
\begin{eqnarray*}
T_{\mu \nu }=\mu ^{2}\rho \delta _{\mu p}\delta _{\nu p}
\end{eqnarray*}%
are fulfilled, with $f$ and $\mu $ \ the same as in the vacuum case and $w$
solution of the $\mu $-\textit{deformed Poisson equation} 
\begin{eqnarray*}
\mathbb{\mu }\Delta w+\nabla \mathbb{\mu }\cdot \nabla w=2f\mathbb{\mu }%
^{2}\rho ,
\end{eqnarray*}%
with $c=1,$ $8\pi G=1$. From the asymptotic flatness condition with $\mu =1$%
, the equation for $w$ reduces to a two-dimensional Poisson equation 
\begin{eqnarray*}
\Delta w=\rho .
\end{eqnarray*}%
It is well known that, if $\rho $ goes to zero fast enough, it is possible
to find non trivial everywhere regular solutions $w$ tending to a constant
value. The function $f$ satisfies the equation 
\begin{eqnarray*}
f\Delta f-\left( \nabla f\right) ^{2}=0\text{,}
\end{eqnarray*}%
and this implies that the function 
\begin{eqnarray*}
\psi =\ln \left\vert f\right\vert
\end{eqnarray*}%
is harmonic. Thus, in order to have everywhere regular spatially
asymptotically flat solutions, $f$ and $\mu $ must be constant functions and
the fluid density $\rho $ must tend to zero fast enough.

However, if we admit $\delta $-like singularity in the $\left( x,y\right) $
plane (\textit{i.e.,} string-like singularity, by taking into account the
third spatial dimension), spatially asymptotically flat vacuum solutions
with $f\neq const$ and $w\neq const$ can exist. In this limiting case in
which $\rho \left( x,y\right) \rightarrow \delta \left( x,y\right) $, the
energy-momentum tensor becomes the one usually employed to describe the
gravitational effects of cosmic strings.

As far as Gamma Ray Bursts sources are concerned, they are usually modelled
with null-like hypersurfaces\footnote{%
Null-like frequently considered as models for astrophysical processes
involving relativistic jets and sudden acceleration of huge quantity of mass 
\cite{BH04}}. In several papers (i.e. \cite{CL96, NAA03}) the study of exact
solutions, describing impulsive gravitational waves, have been posed
starting from a metric tensor of a PP-wave and extended to quadratic
curvature gravity \cite{Ne03, NAA04} and a rough estimation for the
intensity of such waves is provided. Assuming typical parameters for Gamma
Ray Bursts like an energy flux $\backsim 10^{-2}$ $erg\cdot cm^{-2}s^{-1}$,
a temporal extension $\delta t\sim 10s$ and an energy density $\rho
_{0}\approxeq \frac{E}{4\pi z^{2}c\delta t}$ , $z$ being the distance from
the source and $E$ the total energy emitted, the maximal amplitude for a
gravitational wave signal arriving on heart would be of the order $2\pi \rho
_{0}G\delta t^{2}\sim 10^{-38},$ far below the sensitivity of modern
detectors.

\subsection{Zelmanov's and Pirani's criteria}

To check the propagative nature of gravitational fields described by metrics
(\ref{gm}) several covariant criteria have been employed. In the general
case when $f$ and $\mu $ are not constant functions the Zel'manov criterion 
\cite{Za73} is satisfied \cite{CVV04}. Moreover, when $f$ is a constant
function, the only nonvanishing components of the Riemann tensor field
reduce to 
\begin{equation}
R_{txzx}=\frac{w,_{xx}}{2(z-t)^{2}}{,\,}~~\,R_{txzy}=\frac{w,_{xy}}{%
2(z-t)^{2}}{,\,\,}~~R_{tyzy}=\frac{w,_{yy}}{2(z-t)^{2}}  \label{rief}
\end{equation}
which, $w(x,y)$ being a harmonic function, are all harmonic functions of $%
x,y $. As a consequence, the generalized Zel'manov criterion is still
satisfied \cite{CVV04}.

Besides the Zel'manov-Zakharov criterion, the Pirani algebraic criterion is
also satisfied. In light-cone coordinates ($u=\left( z-t\right) /\sqrt{2}%
,\;v=\left( z+t\right) /\sqrt{2}$), where the metrics given by Eq.(\ref{gw})
read 
\begin{equation}
g=2f(dx^{2}+dy^{2})+2dudv+dw~d\ln \left\vert u\right\vert \text{,}
\label{presegw}
\end{equation}
the vector fields $\partial _{u}$ and $\partial _{v}$ are both isotropic.
Moreover, the only non vanishing components of the Riemann tensor are 
\begin{eqnarray*}
R_{uiuj}=\pm \frac{1}{2u^{2}}\partial _{ij}^{2}w
\end{eqnarray*}
and this corresponds to a \textit{type}-\textbf{N} Riemann tensor in the
Petrov classification. From the natural interpretation of the Pirani
criterion \cite{Za73} it follows that the gravitational wave propagates
along the null vector field $\partial _{u}$, that's to say the gravitational
wave (\ref{gw}) propagates along the $z-$axis with the light velocity $c=1$.

\subsection{The energy-momentum pseudo-tensors}

The exact gravitational wave 
\begin{equation}
g=dx^{2}+dy^{2}+dz^{2}-dt^{2}+d\left( w\right) d\left( \ln \left\vert
z-t\right\vert \right) ,  \label{segw}
\end{equation}
given by Eq. (\ref{gw}) for $\mu =1,\,\,f=1/2$ , has the physically
interesting form of a \textit{perturbed Minkowski metric } with $h=dwd\ln
\left\vert z-t\right\vert $ . Moreover, besides being an exact solution of
the Einstein equations, it is a solution of the linearized Einstein
equations on a flat background too: 
\begin{eqnarray*}
\left\{ 
\begin{array}{l}
\eta ^{\mu \nu }\partial _{\mu }\partial _{\nu }h=0 \\ 
\eta ^{\mu \nu }(2h_{\mu \rho ,\nu }-h_{\mu \nu ,\rho })=0\label{wh}%
\end{array}
\right.
\end{eqnarray*}
To study its energy and polarization, the standard tools of the linearized
theory, and in particular the \textit{canonical} energy-momentum
pseudo-tensor, can be used \cite{Di75,We72}.

With $h=d\left( w\right) d\left( \ln \left\vert z-t\right\vert \right) $ the 
$\tau _{0}^{0}$ component of the canonical energy-momentum tensor vanishes
since $h$ has only one index in the plane transversal to the propagation
direction because the components of the tensor $h$ cannot be expressed in
the transverse-traceless gauge.

The non vanishing components of the $4-$momentum density tensor $p^{\mu
}\equiv \tau _{0}^{\mu }$, $\tau ^{\rho \kappa }$ denoting the
Landau-Lifshitz energy-momentum pseudo-tensor \cite{LL76}, are $p^{0}$and $%
p^{3}$. 
\begin{eqnarray*}
\left\{ 
\begin{array}{l}
p^{0}=4\left( t-z\right)
^{-2}[C_{1}(w_{,xx})^{2}+C_{2}(w_{,xy})^{2}]+4\left( t-z\right)
^{-4}C_{3}\nabla {\cdot }[\left\vert \nabla w\right\vert ^{2}\nabla w], \\ 
p^{1}=0,\quad p^{2}=0,\, \\ 
p^{3}=4\left( t-z\right)
^{-2}[C_{1}(w_{,xx})^{2}+C_{2}(w_{,xy})^{2}]+4\left( t-z\right)
^{-4}C_{3}\nabla {\cdot }[\left\vert \nabla w\right\vert ^{2}\nabla w].%
\end{array}
\right.
\end{eqnarray*}
where $C_{i}$ are some positive numerical constants, $\nabla =\left(
\partial _{x},\partial _{y}\right) $ and the harmonicity condition for $w$
has been used \cite{CVV02, CVV04}.

The use of the Bel's superenergy tensor \cite{Be58} 
\begin{eqnarray*}
T^{\alpha \beta \lambda \mu }=\frac{1}{2}\left( R^{\alpha \rho \lambda
\sigma }R_{\,\,\,\,\rho \;\;\sigma }^{\beta \;\;\mu }+\left. ^{\ast
}R\right. ^{\alpha \rho \lambda \sigma }\left. ^{\ast }R\right.
_{\,\,\,\,\rho \;\;\sigma }^{\beta \;\;\mu }\right) ,
\end{eqnarray*}
where the symbol $\ast $ denotes the \textit{volume dual}, leads to the same
result. Indeed, the only non vanishing independent components of the
covariant Riemann tensor \ $R_{\alpha \beta \gamma \delta }=g_{\alpha \rho
}R_{\,\,\,\,\,\beta \gamma \delta }^{\rho }$ are 
\begin{eqnarray*}
R_{1313}=-w,_{11};\,\,\,\,R_{\,1\,323}=-w,_{12};\,\,\,\,\,R_{2323}=-w,_{22}.
\end{eqnarray*}
It follows that the density energy represented by the Bel's scalar 
\begin{eqnarray*}
W=T_{\alpha \beta \lambda \mu }U^{\alpha }U^{\beta }U^{\lambda }U^{\mu },
\end{eqnarray*}
the $U^{\alpha }$'s denoting the components of a time-like unit vector
field, depends on the squares of $w,_{ij}$.

Thus, both the Landau-Lifshitz pseudo-tensor and the Bel superenergy tensor
single out the same physical degrees of freedom. In particular, we can take
the components $h_{tx}$ and $h_{ty}$ as fundamental degrees of freedom for
the gravitational wave (\ref{segw}).\qquad

Since $p^{0}=p^{3},$ these waves move at light velocity, according with the
result obtained by the Pirani criterion.

\subsection{Polarization \textit{\ \ }}

Even more controversial than for the energy and momentum, the definition of
spin or polarization for a theory, such as general relativity, which is
non-linear and possesses a much bigger invariance than just the Poincar\'{e}
one, deserves a careful analysis.

It is well known that the concept of particle together with its degrees of
freedom like the spin may be only introduced for linear theories (for
example for the Yang-Mills theories, which are non linear, it is necessary
to perform a perturbative expansion around the linearized theory). In these
theories, when Poincar\'{e} invariant, the particles are classified in terms
of the eigenvalues of two Casimir operators of the Poincar\'{e} group, $%
P^{2} $ and $W^{2}$ where $P_{\mu }$ are the translation generators and $%
W_{\mu }={\frac{1}{2}}\epsilon _{\mu \nu \rho \sigma }P^{\nu }M^{\rho \sigma
}$ is the \textit{Pauli-Ljubanski polarization vector} with $M^{\mu \nu }$
Lorentz generators. Then, the total angular momentum $J=L+S$ is defined in
terms of the generators $M_{\mu \nu }$ as $J^{i}={\frac{1}{2}}\epsilon
^{0ijk}M_{jk}$ . The generators $P_{\mu }$ and $M_{\mu \nu }$ span the
Poincar\'{e} algebra, $\mathcal{ISO}(3,1)$ 
\begin{equation}
\left\{ 
\begin{array}{l}
\left[ M_{\mu \nu },M_{\rho \sigma }\right] =-i(\eta _{\mu \rho }M_{\nu
\sigma }-\eta _{\mu \sigma }M_{\nu \rho }-\eta _{\nu \rho }M_{\mu \sigma
}+\eta _{\nu \sigma }M_{\mu \rho }) \\ 
\left[ M_{\mu \nu },P_{\rho }\right] =i(\eta _{\nu \rho }P_{\mu }-\eta _{\mu
\rho }P_{\nu }) \\ 
\left[ P_{\mu },P_{\nu }\right] =0.%
\end{array}
\right.  \label{iso31}
\end{equation}

Let us briefly recall a few details about the representation theory of this
algebra. The Pauli-Ljubanski operator is a translational invariant Lorentz
vector, that is $\left[ P_{\mu },W_{\nu }\right] =0$, $\left[ M_{\mu \nu
},W_{\rho }\right] =i(\eta _{\nu \rho }W_{\mu }-\eta _{\mu \rho }W_{\nu })$.
In addition it satisfies the equation 
\begin{equation}
W_{\mu }P^{\mu }=0.  \label{wp}
\end{equation}
The unitary (infinite-dimensional) representations of the Poincar\'{e} group
fall mainly into three different classes:

\begin{itemize}
\item $P^{2}=m^{2}>0$, $W^{2}=-m^{2}s(s+1)$, where $s=0,{\frac{1}{2}},1,...$
denotes the \textit{spin}. From Eq. (\ref{wp}) we deduce that in the rest
frame the zero component of the Pauli-Ljubanski vector vanishes and its
space components are given by $W_{i}={\frac{1}{2}}\epsilon
_{i0jk}P^{0}S^{jk} $ so that $W^{2}=-m^{2}S^{2}$. This representation is
labelled by the mass $m $ and the spin $\emph{s}$.

\item $P^{2}=0$, $W^{2}=0$. In this case $W$ and $P$ are linearly dependent 
\begin{eqnarray*}
W_{\mu }=\lambda P_{\mu };
\end{eqnarray*}
the constant of proportionality is called \textit{helicity} and it is equal
to $\pm \emph{s}$ . The time component of $W$ is $W^{0}=\overrightarrow{P}{%
\cdot }\overrightarrow{J}$, so that 
\begin{eqnarray*}
\lambda =\frac{\overrightarrow{P}{\cdot }\overrightarrow{J}}{P_{0}}
\end{eqnarray*}
which is the definition of helicity for massless particles like photons.

\item $P^{2}=0$, $W^{2}=-\rho ^{2}$, where $\rho $ is a continuous
parameter. This type of representation, which describes particles with zero
rest mass and an infinite number of polarization states labeled by $\rho $,
does not seem to be realized in nature.
\end{itemize}

Let us turn now to the gravitational fields represented by Eq. (\ref{segw}).
As it has been shown, they represent gravitational waves moving at the
velocity of light, that is, in the would be quantised theory, particles with
zero rest mass. Thus, if a classification in terms of Poincar\'{e} group
invariants could be performed, these waves would belong to the class of
unitary (infinite-dimensional) representations of the Poincar\'{e} group
characterized by $P^{2}=0$, $W^{2}=0$. Recall that, in order for such a
classification to be meaningful $P^{2}$ and $W^{2}$ have to be invariants of
the theory. This is not the case for general relativity, unless we restrict
to a subset of transformations selected for example by some physical
criterion or by experimental constraints. For the solutions of the
linearized vacuum Einstein equations the choice of the harmonic gauge does
the job \cite{We72}. There, the residual gauge freedom corresponds to the
sole Lorentz transformations.

The polarization of gravitational fields represented by Eq. (\ref{presegw})
can be estimated by looking at the transformation properties of the two
independent physical components of this metrics under a rotation in the
plane $\left( x,y\right) $ orthogonal to the propagation direction. This
physical components, $h_{tx}$ and $h_{ty}$, have only one index in the $%
\left( x,y\right) $ plane orthogonal to the propagation direction $\partial
_{u}$. Under the infinitesimal rotation $\mathcal{R}$ in the $\left(
x,y\right) -$plane they transform as a vector.

Applied to any vector $\left( v_{1},v_{2}\right) $ the infinitesimal
rotation generator $\mathcal{R}$ , has the effect 
\begin{eqnarray*}
\mathcal{R}v_{1}=v_{2}\,,\,\,\,\,\,\mathcal{R}v_{2}=-v_{1},
\end{eqnarray*}%
from which 
\begin{eqnarray*}
\mathcal{R}^{2}v_{i}=-v_{i}\,\,\,\,\ i=1,2,
\end{eqnarray*}%
so that $i\mathcal{R}$ has the eigenvalues $\pm 1$. Thus, the components of $%
h_{\mu \nu }$ that contribute to the energy correspond to spin-$1$ fields.
The reason why it's commonly believed that spin-1 do not exist is that, in
treating with the linearized theory, solutions are implicitly assumed to be
square integrable: this is not the case for solution\ like metrics (\ref%
{presegw}).

These solutions are interesting for (at least) two reasons. First, they are
asympotically flat (at least with a $\delta $-like singularity) in the plane
transversal to the propagation direction. Second, they are both solutions of
the linearized theory and of the exact theory, so that the spin-$1$ result
cannot be attributed to the first order approximation.

\subsection{Raychaudhuri equation}

A possible approach to the analysis of physical effects of gravitational
waves is based on the Raychaudhuri equation. This approach has the important
advantage to be covariant so that it is not needed anymore to care about the
choice of a coordinate system. The evolution of a beam (congruence) of non
light-like curves (i.e., trajectories of test masses which one would like to
observe) on curved space-time is constrained by the Raychaudhuri equation: 
\begin{align}
\nabla _{A}\theta & =-R_{cb}A^{c}A^{b}+2\omega ^{2}-2\sigma ^{2}-\frac{%
\theta ^{2}}{3}+\nabla _{b}\left( \nabla _{A}A\right) ^{b},  \label{re} \\
\omega _{ab}& \equiv \nabla _{\left[ a\right. }A_{\left. b\right]
},~~~\sigma _{ab}\equiv \nabla _{\left( a\right. }A_{\left. b\right)
}-\theta \left( g_{ab}\pm A_{a}A_{b}\right) /3+\left( \nabla _{A}A\right)
_{a}A_{b}  \label{re2} \\
\nabla _{A}& \equiv A^{c}\nabla _{c},\ \theta \equiv \nabla _{b}A^{b},\
A_{c}A^{c}=\mp 1  \notag
\end{align}
where $R_{cb}$ are the components of the Ricci tensor,\textit{\ } $A^{c}$
denotes the vector field tangent to the curve, the signature of $g$ is $%
(+,-,-,-)$, the upper and the lower signs in Eq. (\ref{re2}) correspond
respectively to the space-like and time-like case. The functions $\theta ,$%
\textit{\ }$\sigma _{ab},$ $\omega _{ab}$ are called the \textit{expansion,} 
\textit{shear} and \textit{twist} of the congruence.

In our case it is easier to perform computations in the \textquotedblright
Peres\ system of coordinates\textquotedblright\ where the metrics read: 
\begin{equation}
ds^{2}=dx^{2}+dy^{2}+\varphi (x,y,u)du^{2}+2dudv,~~~\Delta _{(x,y)}\varphi =0
\label{ppw}
\end{equation}%
and, as a timelike curve describing the motion, it is convenient to choose 
\begin{eqnarray*}
A^{a}=\left( 0,0,\frac{1}{\sqrt{\varphi -2}},\frac{-1}{\sqrt{\varphi -2}}%
\right) ,
\end{eqnarray*}%
so that 
\begin{align}
\theta & =-\frac{\partial _{u}\varphi }{2\left( \sqrt{\varphi -2}\right) ^{3}%
},  \notag \\
\quad 2\omega ^{2}-2\sigma ^{2}& =\frac{\left( \partial _{x}\varphi \right)
^{2}+\left( \partial _{y}\varphi \right) ^{2}}{\left( \varphi -2\right) ^{2}}%
-\frac{\left( \partial _{u}\varphi \right) ^{2}}{3\left( \varphi -2\right)
^{3}},  \notag \\
\overset{.}{\theta }& =-\frac{5\left( \partial _{u}\varphi \right) ^{2}}{%
12\left( \varphi -2\right) ^{3}}+\frac{\left( \partial _{x}\varphi \right)
^{2}+\left( \partial _{y}\varphi \right) ^{2}}{\left( \varphi -2\right) ^{2}}%
,  \label{ra3}
\end{align}%
in which, assuming that $\varphi -2>0$, it is possible to isolate the
physical effects of the spin-$1$ gravitational waves described by Eq. (\ref%
{ra3}). It is clear that, due to the non trivial dependence of $\overset{.}{%
\theta }$ on the transverse coordinates, spin-$1$ gravitational waves have
distinguishing features with respect to spin-$2$ gravitational waves. These
effects are manifest, for example, in a distribution of test particles that,
in the case of spin-$1$ waves will experience a permanent displacement that
is, to say, a memory effects \cite{Ch91}.

\section{Detection of gravitational waves}

The observable effect of a gravitational wave acting on two nearby test
masses in free fall is mathematically described by the Jacobi geodesics
deviation equation. If the separation four-vector between two test masses is 
$Z^{\mu }$ the equation can be written in the following form: 
\begin{equation}
\frac{D^{2}Z^{\mu }}{Ds^{2}}=R_{\nu \rho \sigma }^{\mu }Z^{\rho }u^{\nu
}u^{\sigma }  \label{eq. deviazioni geodetiche}
\end{equation}
where $D$\ denotes covariant derivation, $s$ is the \emph{proper time} along
the reference geodesic, $u^{\nu }=dx^{\nu }/ds$ is the velocity four-vector, 
$R_{\nu \rho \sigma }^{\mu }$ is the Riemann tensor.

It's commonly believed that reasonable sources of gravitational waves are so
far from earth that it's always possible to consider the weak field
approximation of \ Einstein's field equations. In this approximation a
gravitational wave is described by the perturbation $h_{\mu \nu }$ to the
Minkowsky metric. Moreover, according with the standard textbook analysis 
\cite{MTW}, it is possible to choose a TT-gauge and express the geodesics
deviation equation (\ref{eq. deviazioni geodetiche}) in the following form: 
\begin{equation}
\frac{d^{2}X^{i}}{dt^{2}}=\frac{1}{2}\delta _{j}^{i}\frac{\partial
^{2}h_{jk}^{TT}}{\partial t^{2}}X^{k},  \label{TT-g eq del moto}
\end{equation}
where the geodesic deviation four-vector $Z^{\mu }$ has been chosen to be $%
\left( 0,\overrightarrow{X}\right) $ where $\overrightarrow{X}$ denotes the
position vector of one of two particles in the comoving reference frame of
the other. The r.h.s. of equation (\ref{TT-g eq del moto}) appears as an
effective $\emph{Newtonian}$ force acting on test masses. An essential
feature of these equation is the possibility of factorizing the time
dependence (see (\ref{forza}) and (\ref{geo dev - arm sph 2}) later).

Now let's consider Eq. (\ref{eq. deviazioni geodetiche}) for metrics (\ref%
{segw}) in weak field and small velocities ($ds\sim dt$) limits. We get: 
\begin{equation}
\frac{d^{2}X^{i}}{dt^{2}}=\frac{\eta ^{ij}X^{k}}{(z-t)^{2}}\partial
_{k}\partial _{j}w,\quad i=1,2,3\text{,}  \label{dev geo svv}
\end{equation}%
Equations (\ref{TT-g eq del moto}) and (\ref{dev geo svv}) show that both
spin-$1$ and spin-$2$ waves are transversal to the propagation direction.
Equations (\ref{dev geo svv}) are less trivial than Eq.s (\ref{TT-g eq del
moto}). They cannot be integrated to give a general solution. Moreover, the
dependence on the \emph{time variable} cannot be factorized. A reasonable
expectation is a permanent deformation on the initial distribution of test
masses, as in the case of the \emph{Christodoulu memory }\cite{Ch91}, a well
known effect due to the non-linearity of the gravitational field. Similar
effects involved in physical process generated by burst sources are called 
\emph{burst with memory} (BWM) \cite{BT87}.

More in general, Eq. (\ref{eq. deviazioni geodetiche}), for metric (\ref%
{spin1}) and $Z^{\mu }=\left( 0,\overrightarrow{X}\right) ,$ reads: 
\begin{equation}
\frac{d^{2}}{ds^{2}}X^{i}=-g^{ih}X^{j}\partial _{j}\partial _{h}\partial
_{u}\varphi .  \label{js1}
\end{equation}%
The above equation shows that one can have either attraction or repulsion
according to the choice of the function $\varphi \left( x,y,u\right) $ which
is constrained, outside the matter source, only by the condition to be a
harmonic function of $x$ and $y$. For example, the choice $\varphi =\rho
(x,y)\sigma (u)$, with $\rho $ a harmonic function of $x$ and $y$ and $%
\sigma $ a decreasing function of $u$, will give repulsion. This is not
surprising because it is known from QFT that spin-odd bosons generate
repulsion between particles of the same charge, the charge, in this case,
corresponding to the mass.

\subsection{Experimental devices}

In this section the possibility of detection of spin-$1$ gravitational waves
\ from the experimental point of view is considered. Nowadays there are two
kind of ground based instruments able to investigate gravitational waves in
the high frequency band ($1Hz\div 10^{4}Hz$): laser interferometers (IFOs)
and resonant mass detectors \cite{Th95}. In spite of the high complexity of
any experimental device (whether IFOs or resonant antennas), the detection
principle is very simple. It essentially consists in a measurement of the
displacement of test masses described by (\ref{eq. deviazioni geodetiche})
and (\ref{TT-g eq del moto}). In the following we will review some basic
features of the interaction mechanism between such instruments and
gravitational waves.

In the case of IFOs the test masses are suspended mirrors and the
displacement induced by gravitational waves is measured by laser
interferometry. At present time the most sensible IFOs are the two LIGO
(with coherent antenna patterns) and VIRGO (with antenna pattern coherent
with the one of GEO). Despite of their higher sensitivity, IFOs have non
isotropic antenna patterns, \ i.e. their sensitivity strongly depends on the
relative orientation of the incoming wave and the plane of the IFO's arms.
Moreover, a single IFOs cannot perform spin measurements. We will not
consider them further.

In the case of resonant antennas the detector is considered as an elastic
body bathed by gravitational waves. The response of the detector can be
studied by making use of the classical theory of elasticity \cite{LL79}; the
generic infinitesimal mass element constituting the detector can be
considered as a test mass and the relative displacement produced by
gravitational waves is measured via the normal modes of oscillation. At
present, cylindrical detectors (or Weber bars) are worldwide spread.\ They
are generally three meters long and two tons heavy cylindrical objects made
of aluminium. The most important resonant bars are those belonging to the
IGEC network: ALLEGRO, AURIGA, EXPLORER, NAUTILUS, NIOBE \cite{IGEC}. Like
an IFOs a single Weber bar cannot perform spin measurements and have
non-isotropic sensitivity. One could imagine a combined use of several
suitably oriented antennas to obtain some information about spin. At present
time this is not feasible because they are oriented with coherent antenna
patterns to reduce the false alarm probability. Therefore, from the point of
view of spin measurement, the whole array is equivalent to a single bar. In
this contest we will focus our attention on spherical detectors because,
unlike other detectors, in principle they are able to determinate the\
polarization of any incoming gravitational wave (and in a wider sense, to
distinguish between different metric theories of gravitation \cite{Wi93}) so
they seem to be the most natural instruments to investigate the spin-$1$
gravitational waves.

In order to describe the effect of spin-$2$ gravitational waves on an
infinitesimal mass element of the detector located at position $%
\overrightarrow{r}=\left( x^{_{1}},x^{_{2}},x^{_{3}}\right) $ in a reference
frame whose origin is at the center of the detector, it will be useful (also
for later purpose) to denote by $f_{s=2}^{i}(\overrightarrow{r},t)$ the
force in Eq. (\ref{TT-g eq del moto}): 
\begin{equation}
f_{s=2}^{i}(\overrightarrow{r},t)\equiv \frac{1}{2}\delta ^{ij}\frac{%
\partial ^{2}h_{jk}(t)}{\partial t^{2}}x^{_{k}}\text{.}  \label{forza}
\end{equation}%
Factorizing the radial and the angular dependence in $\ f_{s=2}^{i}(%
\overrightarrow{r},t)$ one obtains a result depending on spherical harmonics
with $l=2$ only : 
\begin{equation}
f_{s=2}^{i}(\overrightarrow{r},t)=\frac{\partial }{\partial x^{_{j}}}\sqrt{%
\frac{\pi }{15}}\delta ^{ij}r^{2}\sum_{m}\overset{..}{h}_{m}(t)Y_{2m}\text{.}
\label{geo dev - arm sph 2}
\end{equation}

Let $\overrightarrow{u}(\overrightarrow{r},t)$ be the displacement vector of
an infinitesimal mass element, located at position $\overrightarrow{r}$ with
respect to the center of mass of the initially unperturbed solid with
constant density $\rho $ and Lam\'{e} coefficients $\lambda $ and $\mu $.
When a force $\overrightarrow{f}\left( \overrightarrow{r},t\right) $ is
acting on the body, the induced displacement vector $\overrightarrow{u}(%
\overrightarrow{r},t)$ field is solution of the following system of partial
differential equations\footnote{%
Dissipation terms can be easily included in realistic cases}:

\begin{equation}
\rho \frac{\partial ^{2}\overrightarrow{u}}{\partial t^{2}}=(\lambda +%
{\mu}%
)\nabla (\nabla \cdot \overrightarrow{u})+%
{\mu}%
\nabla ^{2}\overrightarrow{u}+\overrightarrow{f}\left( \overrightarrow{r}%
,t\right) .  \label{eq. corpi elastici}
\end{equation}%
In the following we will describe by external force $\overrightarrow{f}%
\left( \overrightarrow{r},t\right) $ the effects coming from the geodesic
deviations\footnote{%
In the context of resonant antennas $\overrightarrow{f}\left( 
\overrightarrow{r},t\right) $ generally contains two contributions: one from
geodesic deviations \ and one describing the forces between the surface of
the elastic body and other objects eventually matched with it (i.e. resonant
transducers suitably tuned to the natural frequencies of oscillation).},
whose explicit espression is given by (\ref{forza}) or equivalently (\ref%
{geo dev - arm sph 2}). A generic solution of Eq.(\ref{eq. corpi elastici})\
is expressed as linear combination of eigenfunctions, with time dependence
appearing in the coefficients $a_{m}(t)$ only: 
\begin{eqnarray*}
\overrightarrow{u}(\overrightarrow{r},t)=\sum_{m}a_{m}(t)\overrightarrow{u}%
_{m}(\overrightarrow{r}),
\end{eqnarray*}%
where $\overrightarrow{u}_{m}(\overrightarrow{r})$ is eigenfunction of the
equation

\begin{equation}
-\rho \omega _{m}^{2}\overrightarrow{u}_{m}=(\lambda +%
{\mu}%
)\nabla (\nabla \cdot \overrightarrow{u}_{m})+%
{\mu}%
\nabla ^{2}\overrightarrow{u}_{m}  \label{eq. autoval}
\end{equation}%
and describes a free oscillation with frequency $\omega _{m}$. This model
applies to any resonant mass detector once we impose the boundary conditions
determined by the detector's shape, i.e. to cylindrical antennas in three
dimensions \cite{PW76} or in one dimensional approximation \cite{Pi93}, to
spherical detectors \cite{CLO95,CFFLO98,Lo95}.

The boundary conditions for sphere's surface free from stress and strain are
expressed by the following relation: 
\begin{equation}
\sigma _{ij}n_{j}=0\text{ at }r:=\left\vert \overrightarrow{r}\right\vert =R,
\label{cond contorno}
\end{equation}%
$R$ being the radius of the sphere, $n_{j}$ the components of the outgoing
surface normal vector, $\sigma _{ij}=\lambda u_{kk}\delta _{ij}+2\mu u_{ij}$
the \emph{stress tensor} and $u_{ij}=\frac{1}{2}\left(
u_{i},_{j}+u_{j},_{i}\right) $ the \emph{strain tensor}. Then, the
time-dependent normal mode amplitude $a_{m}(t)$ is solution of a driven
harmonic oscillator equation:

\begin{equation}
\overset{..}{a}_{m}(t)+\omega _{m}^{2}a_{m}(t)=\frac{1}{\rho N_{m}}\dint 
\overrightarrow{u}_{m}(\overrightarrow{r})\cdot \overrightarrow{f}(%
\overrightarrow{r},t)d^{3}r,  \label{oscill arm forz}
\end{equation}%
with $N_{m}$ normalization constants. All the interaction with external
world (only gravitation in this case) can be included in the r.h.s. as
additional terms of effective force acting on every normal mode. With some
straightforward calculations Eq. (\ref{eq. autoval}) with boundary
conditions (\ref{cond contorno}) can be solved giving the following
solutions : 
\begin{equation}
\overrightarrow{u}(\overrightarrow{r})=\frac{C_{0}}{k_{irr}^{2}}\nabla
\varphi (\overrightarrow{r})+\frac{iC_{1}}{k_{div-free}}\overrightarrow{L}%
\psi (\overrightarrow{r})+\frac{iC_{2}}{k_{div-free}^{2}}\nabla \times 
\overrightarrow{L}\psi (\overrightarrow{r})\text{,}
\label{modi normali sfera}
\end{equation}%
with wave numbers $k_{irr}=\rho \omega ^{2}/\lambda +2\mu $, $%
k_{div-free}=\rho \omega ^{2}/\mu $, where $\varphi (\overrightarrow{r})$
and $\psi (\overrightarrow{r})$ are scalar functions solutions of Helmotz's
equation, $\overrightarrow{L}$ is the angular momentum operator $%
\overrightarrow{L}=-i\overrightarrow{r}\times \nabla ,$ $C_{0},$ $C_{1}$ and 
$C_{2}$ are constants, their numerical value depending on the boundary
conditions (\ref{cond contorno}) in the specific case. To get regular
solutions in $r=0$ the scalar functions $\varphi (\overrightarrow{r})$ and $%
\psi (\overrightarrow{r})$ must take the form: 
\begin{eqnarray*}
\varphi (\overrightarrow{r})=j_{l}(qr)Y_{lm}(\overrightarrow{n})\text{,}%
\quad \psi (\overrightarrow{r})=j_{l}(kr)Y_{lm}(\overrightarrow{n})\text{,}
\end{eqnarray*}%
where $j_{l}$ is a \emph{spherical Bessel function} and $Y_{lm}(%
\overrightarrow{n})$ a\emph{\ spherical harmonic}.

Normal modes of oscillation can be divided in two family: \textit{Thoroidal
modes} ($C_{0}=C_{2}=0$) and \textit{Spheroidal modes} ($C_{1}=0$). The
latter can be expressed in the form: 
\begin{equation}
\overrightarrow{u}_{nlm}(\overrightarrow{r})=A_{nl}(r)Y_{lm}(\overrightarrow{%
n})\overrightarrow{n}-iB_{nl}(r)\overrightarrow{n}\times \overrightarrow{L}%
Y_{lm}(\overrightarrow{n})\text{,}  \label{modi sferoidali}
\end{equation}%
with $A_{nl}(r)$ and $B_{nl}(r)$ combinations of Bessel functions.

Let's go back to the \emph{effective force} (\ref{oscill arm forz}) acting
on every normal mode. If we chose for $\overrightarrow{f}(\overrightarrow{r}%
,t)$ the expression (\ref{geo dev - arm sph 2}), the only non-zero integrals
on the r.h.s. of Eq. (\ref{oscill arm forz}) will be those containing the
eigenfunctions $\overrightarrow{u}_{nlm}$ with spherical harmonics $%
Y_{2m}(\theta ,\varphi )$: 
\begin{eqnarray*}
\overrightarrow{u}_{lm}=[\alpha _{l}(r)\overrightarrow{n}+\beta
_{l}(r)R\nabla ]Y_{lm}(\theta ,\varphi )\text{.}
\end{eqnarray*}%
Functions $\alpha _{l}(r)$ and $\beta _{l}(r)$ are combinations of spherical
Bessel function of order $2$ and determinate the motion in radial and
tangential direction respectively. According with this standard analysis,
only few modes of the spheres, the five quadrupolar modes ($l=2$), will be
coupled to gravity. Moreover, several studies \cite{MJ97} show that a finite
number of resonators opportunely tuned to this mode's frequency, will
suffice to completely solve the problem of deconvolution of the signal
revealed by spheres.

Suppressing the index $l$ we can write the r.h.s. of Eq. (\ref{oscill arm
forz}) as: 
\begin{equation}
F_{m}(t)=\int_{sphere}\overrightarrow{u}_{m}\cdot \overrightarrow{f}%
_{s=2}d^{3}r=\frac{1}{2}\overset{..}{h}_{m}(t)\gamma MR\text{,}
\label{forza effettiva}
\end{equation}%
the constant $\gamma $ depending on the physical properties of the elastic
medium.

In the case of \ spin-$1$ gravitational waves relation (\ref{geo dev - arm
sph 2}) cannot be used. If we want to take into account the whole
interaction of spin-$1$ waves with spherical detectors, we will be compelled
to solve the integral in the r.h.s. of Eq. (\ref{oscill arm forz}) making
use of geodesic deviation (\ref{dev geo svv}) and (\ref{js1}) whose r.h.s.
will be denoted by $f_{s=1}^{i}(\overrightarrow{r},t)$. Let us consider the
most simple case choosing the $l=0$ normal mode (\ref{modi normali sfera}).
For $l=0$ thoroidal modes are absent while the non vanishing contribution to
spheroidal modes (\ref{modi sferoidali}) comes from the term $%
A_{nl}(r)Y_{lm}(\overrightarrow{n})\overrightarrow{n}$. Even if it cannot be
explicitly evaluated, there are no reasons for integral on the r.h.s. of (%
\ref{oscill arm forz}) to vanish 
\begin{eqnarray*}
F_{0}(t)=\int_{sphere}\overrightarrow{u}_{n0}\cdot \overrightarrow{f}%
_{s=1}d^{3}r\neq 0
\end{eqnarray*}%
This computation, which heavily depends on the choice of the harmonic
function $w$, does not lead to a general solution as in the standard spin-$2$%
\ case, but stresses the interaction between an acoustic detector and a spin-%
$1$ gravitational wave. Moreover, since $F_{0}$ is non vanishing, spherical
modes with $l=0$ are activated \textit{even} in General Relativity too.
Thus, the standard identification of normal mode's index $l$ with the spin
component of the driving gravitational wave is not ensured. A further step
in this direction would be to test the coupling between normal modes and
spin components in more general cases.

\section{Conclusions}

It is still deep-seated the belief that spin-$1$ gravitational waves cannot
be present, not only in General Relativity but in every metric theory of
gravitation. It has been shown that this erroneous belief derives from the
implicit assumption of considering only square integrable solutions of the
linearized Einstein equations: indeed it is not true that it is always
possible to reduce to TT-gauge and to remove all the \textit{non spin-}$%
\mathit{2}$\textit{\ components} by a gauge transformation.

Once we accept that gravitational waves may have spin-$1$ and\ may be
emitted by reasonable sources, it becomes important to define the
experimental conditions necessary to observe their spin. The first concrete
\ possibility could be the use of spherical detectors.. Due to their
resonant spectrum, described by Eq. (\ref{modi normali sfera}), spherical
detectors are the ideal instruments for studying the polarization. According
to the standard theory, spherical devices are prepared to observe only spin-$%
0$ and spin-$2$ waves. In such an analysis the $(x,y)-$coordinates
dependence is not taken into account. Clearly it will be difficult to
detected spin-$1$ gravitational waves with instruments having dimensions
smaller than typical length scale of spatial variation of the waves. This
does not implies that spin-$1$ waves do not exist at all or that it is not
possible to conceive new experimental apparatus capable to measure their
spin.


\begin{thebibliography}{99}
\bibitem{IGEC} P. Astone et al., \textit{Methods and results of the IGEC
search for burst gravitational waves in the years 1997--2000,} Phys. Rev. D%
\textbf{\ 68,} (2003) 022001.

\bibitem{AL92} B.N. Aliev and A.N. Leznov, \textit{Exact solutions of the
vacuum Einstein's equations allowing for two noncommutative killing vectors
(Type G}$_{\mathit{2}}$\textit{II of Petrov classification)} , J. Math.
Phys. \textbf{33}, (1992) 2567 .

\bibitem{Be58} L. Bel, Compt. Rend. Acad. Sci. Colon\textit{. }\textbf{247}
(1958) 1094\textit{.}

\bibitem{BGPPR94} J. Barret, G.W. Gibbons, M.J. Perry, C. N. Pope and P.
Ruback, \textit{Kleinian geometry and} \textit{N = 2 superstring}. Int. J.
Mod. Phys. A \textbf{9}, (1994) 1457.

\bibitem{BH04} C. Barrabes, P.A. Hogan, \textit{Singular Null Hypersurfaces
in General Relativity: Light-Like Signals from Violent Astrophysical Events, 
}World Scientific (2004).

\bibitem{BK70} V.A. Belinsky and I.M. Khalatnikov\textit{, General solution
of the gravitational equations with aphysical singularities}, Sov. Phys.
JETP \textbf{30}, (1970) 6.

\bibitem{BZ78} V. A. Belinsky and V.E. Zakharov, \textit{Integration of the
Einstein equations by means of the inverse scattering problem technique and
construction of the exact soliton solutions}. Sov. Phys. JETP \textbf{48},
(1978) 6;\textbf{\ }\textit{Stationary gravitational solitons with axial
symmetry.}\textbf{\textit{\ }}\textit{Sov. Phys. JETP}\textbf{\ 50}, (1979)
1.

\bibitem{BT87} V.B. Braginsky, K.S.Thorne, \textit{Gravitational-wave bursts
with memory and experimental prospects}, Nature (London) \textbf{327},
(1987) 123.

\bibitem{CFV04a} D. Catalano Ferraioli, A.M. Vinogradov, \textit{Ricci-flat
4-metrics with bidimensional null orbits, Part I. General aspects and
nonabelian case}.preprint DIPS- 7/2004

\bibitem{CFV04b} D. Catalano Ferraioli, A.M. Vinogradov, \textit{Ricci-flat
4-metrics with bidimensional null orbits, Part II. Abelian case}. Preprint
DIPS- 8/2004

\bibitem{Ch91} D. Christodoulou, \textit{Nonlinear Nature of Gravitaion and
Gravitational-Wave Experiments, }Phys. Rev. Lett. \textbf{67}, (1991) 1486.

\bibitem{Ch98} F.J. Chinea, \textit{New first integral for twisting type-N
vacuum gravitational fields with two non-commuting Killing vectors}, Class.
Quantum Grav. \textbf{15}, (1998) 367.

\bibitem{CL96} M. Campanelli, C. O. Lousto, \textit{Exact Graviattional
Shock Wave Solution of Hiher order Theories, }Phys.Rev. D \textbf{54} (1996)
3854.

\bibitem{CV04} F. Canfora, G. Vilasi, \textit{Spin-1 gravitational waves and
their natural sources,} Phys. Lett.\textit{\ }B, \textbf{585} (2004) 193.

\bibitem{CVV02} F. Canfora, G. Vilasi, P. Vitale, \textit{Nonlinear
gravitational waves and their polarization,} Phys. Lett.\textit{\ }B, 
\textbf{545} (2002) 373.

\bibitem{CVV04} F. Canfora, G. Vilasi, P. Vitale, \textit{Spin-1
gravitational waves},\textit{\ }Int. J. Mod. Phys. B\textbf{\ 18} (2004) 527.

\bibitem{CLO95} E. Coccia, J.A. Lobo, and J.A. Ortega, \textit{Proposed
gravitational wave observatory based on solid elastic spheres}, Phys. Rev. 
\textbf{D }52, (1995) 3735.

\bibitem{CFFLO98} E. Coccia, V. Fafone, G. Frossati, J.A. Lobo, and J.A.
Ortega, \textit{Hollow sphere as a detector of gravitational radiation},
Phys. Rev. \textbf{D} 57, (1998) 2051.

\bibitem{Di75} P.A.M. Dirac, \textit{General Theory of Relativity}, (Wiley,
N.Y. 1975).

\bibitem{ER37} A. Einstein and N. Rosen, \textit{On gravitational waves}, J.
Franklin Inst. \textbf{223}, (1937) 43.

\bibitem{Ge72} R. Geroch, \textit{A Method for Generating New Solutions of
Einstein's Equation. II},\textit{\ }J. Math. Phys. \textbf{13}, (1972) 394.

\bibitem{Ha88} M. Hallisoy,\textit{\ Studies in space-times admitting two
spacelike Killing vectors, }J. Math. Phys. \textbf{29}, (1988) 320.

\bibitem{HH83} J.B. Hartle and S.W. Hawking, \textit{Wave function of the
universe}. Phys. Rev. D \textbf{28}, (1983) 2960.

\bibitem{Ko58} A.S. Kompaneyets, \textit{Strong Gravitational Waves in free
space}, Sov. Phys. JETP \textbf{7}, (1958) 659.

\bibitem{LL76} L.D.Landau, E.M. Lifshitz, \textit{Teorie du Champ,} (Mir,
1976).

\bibitem{LL79} L.D. Landau, E.M. Lifshitz , \textit{Teorie de l'\'{e}lasticit%
\'{e}}, (Mir, 1979).

\bibitem{Lo95} J.A. Lobo, \textit{What can we learn about GW Physics with an
elastic spherical antenna?, }Phys. Rev. D \textbf{52}, (1995) 591.

\bibitem{MJ97} S.M. Merkowitz, W.W. Johnson, \textit{The TIGA technique for
detecting gravitational waves with a spherical antenna, }Phys. Rev.\textbf{\ 
}D \textbf{56}, (1997) 7513.

\bibitem{MTW} C.W. Misner, K.S. Thorne and J.A. Wheeler, \textit{Gravitation,%
} (W.H. Freeman and Co., San Francisco, 1973).

\bibitem{NAA03} E.C. de Rey Neto, J.C.N. de Araujo and O.D. Aguiar, \textit{%
A gravitational shock wave generated by a beam of null matter in quadratic
gravity,} Class. Quantum Grav. \textbf{20} (2003) 1479.

\bibitem{Ne03} E.C. de Rey Neto, \textit{Geodesic deviation in pp-wave
spacetimes of quadratic curvature gravity}, Phys.Rev. D \textbf{68}, (2003)
124013.

\bibitem{NAA04} E.C. de Rey Neto, J.C.N. de Araujo, O.D. Aguiar, \textit{%
Wave polarizations for a beam-like gravitational wave in quadratic curvature
gravity}, Class.Quant.Grav. \textbf{21} (2004) S541.

\bibitem{OV91} H. Ooguri and C. Vafa, \textit{N = 2 heterotic strings.}
Nucl. Phys. B \textbf{367}, (1991) 83.

\bibitem{Pe59} A. Peres, \textit{Some Gravitational Waves}, Phys. Rev. Lett. 
\textbf{3}, (1959) 571.

\bibitem{Pe60} A. Peres, \textit{Null Electromagnetic Fields in General
Relativity Theory}, Phys. Rev\textit{.} \textbf{118}, (1960) 1105.

\bibitem{Pe69} A.Z.Petrov, \textit{Einstein spaces, }(Pergamon Press,\textit{%
\ } New York, 1969).

\bibitem{Pi93} G. Pizzella, \textit{Fisica sperimentale del campo
gravitazionale}, (Nuova Italia Scientifica, Roma 1993).

\bibitem{PW76} H.J. Paik, R.V. Wagoner, \textit{Calculation of the
absorption cross section of a cylindrical gravitational-wave antenna}, Phys.
Rew. D \textbf{13,} (1976) 2694.

\bibitem{Ro54} N. Rosen, Bull. Res. Council Isr. \textbf{3}, (1954) 328.

\bibitem{Sa03} M. Sazhin et al., \textit{CSL-1: a chance projection effect
or serendipitous discovery of a gravitational lens induced by a cosmic
string?}, MINRAS, \textbf{343}, (2003) 353.

\bibitem{SKMHH03} H. Stephani, D. Kramer, M.MacCallum, C. Honselaers,
E.Herlt, \textit{Exact solutions of Einstein field equations }(Second
Edition, Cambridge University Press, Cambridge 2003).

\bibitem{SV00} G. Sparano and G. Vilasi, \textit{Noncommutative
integrability and recursion operators}, J. Geom. Phys. \textbf{36}, (2000)
270.

\bibitem{SVV01} G. Sparano, G. Vilasi, A.M. Vinogradov, \textit{%
Gravitational fields with a non-Abelian, bidimensional Lie algebra of
symmetries}, Phys. Lett. B\textbf{\ 513}, (2001) 142.

\bibitem{SVV02a} G. Sparano, G. Vilasi, A.M. Vinogradov, \textit{Vacuum
Einstein metrics with bidimensional Killing leaves. I. Local aspects}, Diff.
Geom. Appl. \textbf{16}, (2002) 95.

\bibitem{SVV02b} G. Sparano, G. Vilasi, A.M. Vinogradov, \textit{Vacuum
Einstein metrics with bidimensional Killing leaves. II. Global aspects},
Diff. Geom. Appl. \textbf{17}, (2002) 1.

\bibitem{Th92} K.S. Thorne, \textit{Gravitational-wave bursts with memory:\
The Christodoulou effect, }Phys. Rev. D \textbf{45}, (1992) 520.

\bibitem{Th95} K.S. Thorne, \textit{Gravitational Waves}, [gr-qc/9506086].

\bibitem{We72} S. Weinberg, \textit{Gravitation and Cosmology} (J. Wiley \&
Sons, N. Y., 1972).

\bibitem{Wi93} C.M. Will, \textit{Theory and Experiment in Gravitational
Physics}, (Cambridge University Press, 1993).

\bibitem{Za73} V.D. Zakharov, \textit{Gravitational waves in Einstein's
theory,} (Halsted Press, N.Y.1973).
\end{thebibliography}
\end{document}